\documentclass[Physsubmission, Phys]{SciPost}
\hypersetup{
    colorlinks,
    linkcolor={red!50!black},
    citecolor={blue!50!black},
    urlcolor={blue!80!black}
}
\usepackage{cite}
\usepackage{slashed}
\usepackage[bitstream-charter]{mathdesign}
\def\mathswitch#1{\relax\ifmmode#1\else$#1$\fi}
\def\mathswitchr#1{\relax\ifmmode{\mathrm{#1}}\else$\mathrm{#1}$\fi}
\newcommand{\PW}{\mathswitchr W}
\newcommand{\PZ}{\mathswitchr Z}

\newcommand{\Pt}{\mathswitchr t}

\newcommand{\MW}{\mathswitch {M_\PW}}
\newcommand{\MZ}{\mathswitch {M_\PZ}}

\newcommand{\mw}{\mathswitch {\overline{M}_\PW}}
\newcommand{\mz}{\mathswitch {\overline{M}_\PZ}}

\newcommand{\as}{\alpha_{\mathrm s}}
\newcommand{\seff}[1]{\sin^2\theta_{\rm eff}^{#1}}
\newcommand{\msbar}{$\overline{\mbox{MS}}$}

\newcommand{\gev}{\,\, \mathrm{GeV}}

\newcommand{\SLASH}[2]{\makebox[#2ex][l]{$#1$}/}

\newcommand{\pslash}{\SLASH{p}{.2}}

\newcommand{\OO}{{\mathcal O}}

\newcommand{\mycaption}[1]{\caption{\sl #1}}

\DeclareSymbolFont{usualmathcal}{OMS}{cmsy}{m}{n}
\DeclareSymbolFontAlphabet{\mathcal}{usualmathcal}

\begin{document}

\begin{center}{\Large \textbf{
Leading fermionic three-loop corrections to electroweak precision observables\\
}}\end{center}

% TODO: write the author list here. Use initials + surname format.
% Separate subsequent authors by a comma, omit comma at the end of the list.
% Mark the corresponding author with a superscript *.
\begin{center}
Lisong Chen,
Ayres Freitas
\end{center}

\begin{center}
 Pittsburgh Particle-physics Astro-physics \& Cosmology Center (PITT-PACC)\\
Department of Physics \& Astronomy, University of Pittsburgh, Pittsburgh, PA 15260, USA
afreitas@pitt.edu\\
lic114@pitt.edu
\end{center}

\begin{center}
\today
\end{center}

% For convenience during refereeing (optional),
% you can turn on line numbers by uncommenting the next line:
%\linenumbers

\definecolor{palegray}{gray}{0.95}
\begin{center}
\colorbox{palegray}{
  \begin{tabular}{rr}
  \begin{minipage}{0.1\textwidth}
    \includegraphics[width=35mm]{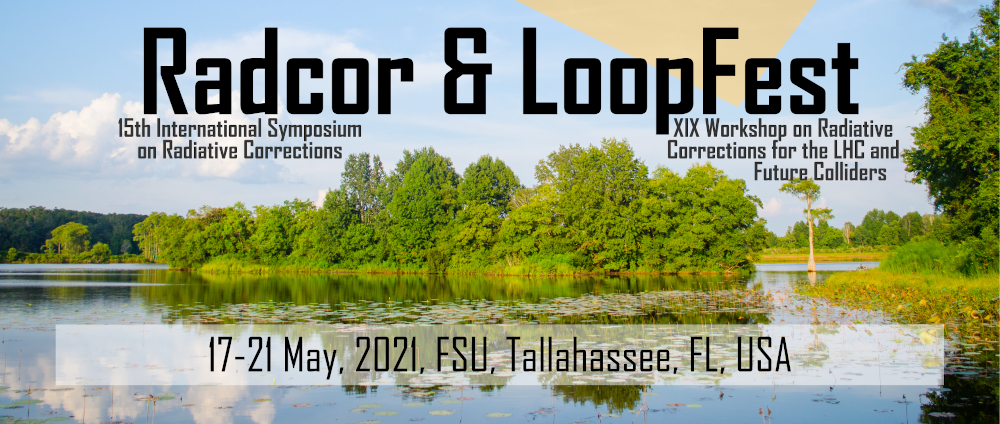}
  \end{minipage}
  &
  \begin{minipage}{0.85\textwidth}
    \begin{center}
    {\it 15th International Symposium on Radiative Corrections: \\Applications of Quantum Field Theory to Phenomenology,}\\
    {\it FSU, Tallahasse, FL, USA, 17-21 May 2021} \\
    \doi{10.21468/SciPostPhysProc.?}\\
    \end{center}
  \end{minipage}
\end{tabular}
}
\end{center}

\section*{Abstract}
{\bf
% TODO: write your abstract here.
%Electroweak precision observables (EWPOs), as the testbed of the Standard Model, will be scrutinized at future high-luminosity $e^+e^-$ colliders with substantial upgrading precision. One can only fully appreciate these high-precision measurements with %accurate theoretical predictions whose uncertainties are under well-controlled. In this 

In this proceeding, we highlight the computation of leading fermionic three-loop corrections to electroweak precision observables (EWPOs) accomplished recently. We summarize the numerical analysis and provide an outlook.
}

\vspace{10pt}
\noindent\rule{\textwidth}{1pt}
\tableofcontents\thispagestyle{fancy}
\noindent\rule{\textwidth}{1pt}
\vspace{10pt}

\section{Introduction}
\label{sec:intro}
% TODO: write your article here.
The electroweak precision observables (EWPOs) are a group of quantities associated with the properties of the Z and W bosons. They can be obtained from measurements of processes mediated by W and Z bosons, where the experimentally irreducible background has been carefully removed. The EWPOs, as one of the most crucial testbeds of the Standard Model (SM),  played a key role in the physics program of LEP and SLC and they will be further
scrutinized at future high-luminosity $e^+e^-$ colliders, such as  FCC-ee, ILC, CLIC, and CEPC, with substantially improved precision. One can only fully take advantage of these high-precision measurements with accurate theoretical predictions whose uncertainties are well-controlled. 
%This means the future advances in the interpretation of EWPOs' measurements at future colliders would lean on the 
 The latter require calculations of multi-loop radiative corrections together with the better knowledge of theory input parameters. Up till now, the theoretical predictions of the EWPOs, such as $(i)$ the W-boson mass $M_W$, $(ii)$ the partial widths of the Z-boson $\Gamma_f$, and $(iii)$ the effective weak mixing angle $\sin^2\theta_{eff}^f$, have been known up to full two-loop level\cite{qcd2,mwshort,mwlong,mw,mwtot,swlept,swlept2,swbb,gz,zbos}, and partial three- and four-loop level contributions given by top Yukawa coupling enhancement\cite{qcd3,mt6,qcd4} within the SM. All these corrections amount to predictions with theoretical uncertainties being safely below the current experimental precision (see Ref~\cite{rev1,rev2,pdg} for detailed reviews). Yet the expected precision of future $e^+e^-$ colliders impose the need of computing three and four-loop corrections at full EW $\mathcal{O}(\alpha^3)$ and mixed EW-QCD $\mathcal{O}(\alpha^2\alpha_s)$ and $\mathcal{O}(\alpha \alpha_s^2)$. 
In this proceeding, we survey the recently accomplished calculations of leading fermionic three-loop corrections to EWPOs at full EW $\mathcal{O}(\alpha^3)$ and mixed EW-QCD $\mathcal{O}(\alpha^2\alpha_s)$, where ``leading fermionic'' refers to the maximal number of closed fermionic loops at given orders. In sec.~\ref{sec:reno}, we introduce the renormalization procedures for cases with and without QCD contributions. Sec.~\ref{sec:tech} highlights the technical aspects including the derivative and evaluation of the master integral (MI) and the computer algebra tools we used. One can find numerical results and shed light on future projections thereby in sec.~\ref{sec:num} and the Conclusion, respectively. 

\section{Renormalization}\label{sec:reno}
\subsection{Renormalization Schemes}\label{schemes}
We adopted the on-shell (OS) renormalization scheme for all electroweak radiative corrections. For corrections involving QCD, such as the case of leading fermionic three-loop at $\mathcal{O}(\alpha^2\alpha_s)$, where the top-quark mass receives radiative corrections from gluon exchange, we use OS scheme and modified minimal subtraction scheme ($\overline{\text{MS}}$) alternately to describe the renormalized top-quark mass. The reason for using both schemes is the following:  the OS top mass definition is subject to the renormalon ambiguity from which the $\overline{\text{MS}}$ top-quark mass prescription is exempt. The $\overline{\text{MS}}$ top-quark mass prescription is thus preferable in practical calculations, yet an extra step is required to relate the {\msbar} value to an observable. These two schemes are related by a finite function, which has been carried out up to four-loop level\cite{osmsbar}. %One should also keep in mind 
The results carried out in both schemes after summing up all orders in perturbation theory should converge up to non-perturbative effects, and our numerical comparison between two schemes will reveal an inkling of it (see \ref{sec:num}).\\
In the OS scheme, the physical mass of the massive unstable particle is defined to be the real part of the complex pole of the propagator, while the width is proportional to the imaginary part of the pole as follows,
\begin{equation}
s_0\equiv \overline{M}^2-i \overline{M}\overline{\Gamma}\label{eq:pole},
\end{equation} 
where $\overline{M}$ is the renormalized mass defined to be on-shell and the $\overline{\Gamma}$ is the width.\footnote{The mass and width defined here are theoretically well-defined and gauge-invariant \cite{zpole}, but the experimental mass and width $M$, $\Gamma$ used, are related to $\overline{M}$, $\overline{\Gamma}$ by the relations 
$\overline{M} = M\big/\sqrt{1+\Gamma^2/M^2},$
$\overline{\Gamma} = \Gamma\big/\sqrt{1+\Gamma^2/M^2}$ \cite{mrel}.}.
For a massive gauge boson, by requiring the inverse of Dyson re-summed two-point function to be zero at the pole as
\begin{equation}
D(s)=Z(s-\overline{M}^2) - \delta \overline{M}^2 Z+\Sigma(s)\lvert_{s=s_0}\equiv0, 
\end{equation}
we get the renormalization conditions
\begin{align}
\delta \overline{M}^2&=Z^{-1}\Re \Sigma(\overline{M}^2-i \overline{M}\overline{\Gamma})\label{rcb1}\\
\overline{\Gamma}&=\frac{\Im\Sigma(\overline{M}^2-i \overline{M}\overline{\Gamma})}{ZM}\label{rcb2}
\end{align}
When deriving the renormalization condition for the Z boson, a more subtle complexity emerges from taking $\gamma-Z$ mixing effect into account (see detailed discussion in Ref.~\cite{Chen:2020xzx}).
The renormalization conditions for massive fermions, akin to massive gauge boson cases, can also be obtained through $D_{\psi}(\pslash)\lvert_{p^2=M_{\psi}^2-iM_{\psi}\Gamma_{\psi}}=0$, where $D_{\psi}$ is the inverse of fermion two-point function written as
\begin{equation}
\begin{split}\label{eq:topp}
D_{\psi}(p)&=Z_{\psi}(\slashed{p}-M_\psi)+\Sigma_{\psi}(p^2)-Z_{\psi}\delta M_{\psi}.
\end{split}
\end{equation}
Hence we get the top-quark mass counterterm and width as
\begin{equation}
\begin{split}
\delta M_\psi u(p) &=Z_{\psi}^{-1}\Re{\Sigma_{\psi}(\slashed{p})u(p)\rvert_{p^2=M_{\psi}^2-iM_{\psi}\Gamma_{\psi}}}\\
\Gamma_\psi u(p)&=Z_{\psi}^{-1}2 \Im \Sigma_{\psi}(\slashed{p})u(p) \rvert_{p^2=M_{\psi}^2-iM_{\psi}\Gamma_{\psi}}.
\end{split}
\end{equation}

By recursively applying the renormalization conditions eq.~\eqref{rcb1} eq.~\eqref{rcb2}, we can obtain widths and mass counterterms in terms of 1-PI self-energies up to arbitrary orders (see explicit expressions in Ref.~\cite{Chen:2020xzx}). Since all EWPOs we want to compute are extracted from processes where the massive gauge bosons appear to be intermediate states only, the final results should be independent of field strength renormalization constants (we have checked it explicitly in our calculations). It is thus safe to set Z to be 1 in our cases.

\begin{figure}[tb]
\centering
\begin{equation}
\begin{split}
\Sigma_{V_1V_2(\alpha^3)} & =
\raisebox{-20pt} {\includegraphics[width=0.5\linewidth]{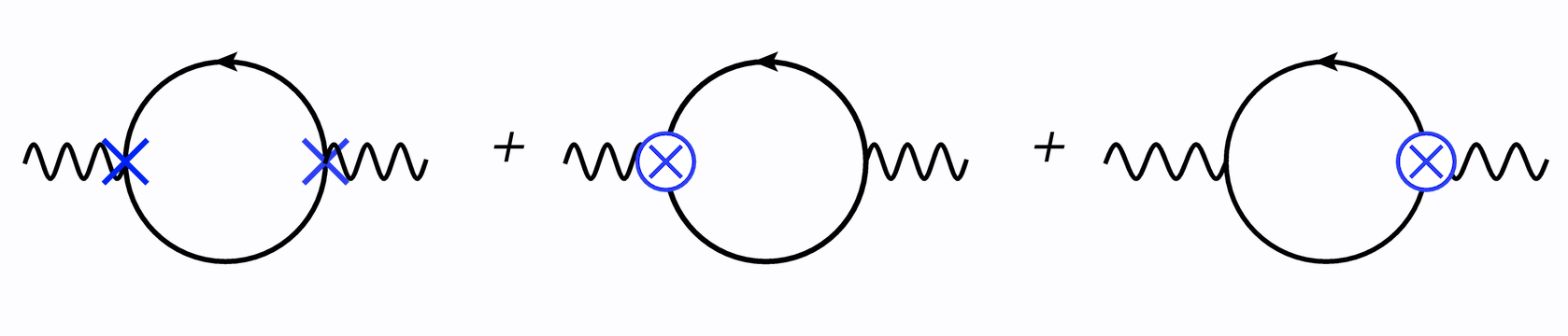}}\\
\Sigma_{V_1V_2(\as\alpha^2)} & = 
\raisebox{-20pt} {\includegraphics[width=0.75\linewidth]{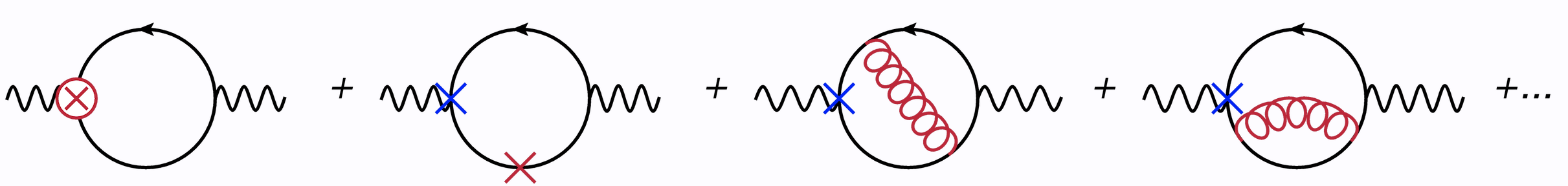}}
\end{split}  \notag
\end{equation}
\vspace{-2em}
\mycaption{Diagramatic 1-PI leading fermionic self-energy functions at
different orders. $V_1$ and $V_2$ denote the possible different in- and outgoing gauge bosons. Vertices "$\otimes$" and "$\times$" indicate the counterterms at the loop order $\mathcal{O}(\as\alpha)$ or $\OO(\alpha^2)$,  and $\mathcal{O}(\alpha)$ or $\mathcal{O}(\as)$, distinguished by red (with QCD) and blue (without QCD), respectively. 
\label{fig:sesub}}
\end{figure}

In the {\msbar} scheme, the mass counterterm is meant to subtract the ultraviolet divergent piece along with constants $\log(4\pi)$ and $\gamma_E$. At one-loop QCD level, it is 
\begin{equation}
\delta m_{t}=-\frac{3 C_F g_s^2}{16 \pi^2}\Bigl(\frac{1}{\epsilon}+\log{4\pi}-\gamma_E\Bigr)\,m_t(\mu).
\end{equation}
And it relates to the OS top-quark mass by
\begin{equation}
\frac{M_t}{m_t}=1+\frac{\as C_F}{4\pi} \Bigl(3\log{\frac{M^2_t}{\mu^2}-4}\Bigr)+\mathcal{O}(\as^2). \label{massrel}
\end{equation}
at one-loop level in QCD.
Moreover, the renormalized weak mixing angle is defined by demanding that the relation $\sin^2\theta_W=1-\mw^2/\mz^2$ holds to all orders. 
The electromagnetic charge, as a fundamental parameter, is renormalized to the coupling strength in Thompson scattering. 
Due to the non-perturbative contribution of light-quark fermionic loops at zero momentum in the Thompson limit, this contribution, parametrized as $\Delta\alpha_{had}$, is usually extracted from measurements of $e^{+}e^{-}\to hadrons$\cite{dahad}.

\subsection{EWPOs definitions}\label{defewpos}

\subsubsection{Fermi constant $G_{\mu}$}\label{fermig}
The Fermi constant can be precisely obtained from muon decay. In the SM, it
is defined as
\begin{equation}
G_{\mu}=\frac{\pi\alpha}{\sqrt{2}s^2_w\mw^2}(1+\Delta r)\label{gmu},
\end{equation}
where all QED contributions have already been taken into account in the determiation of $G_{\mu}$ from the muon lifetime. Here $\Delta r$ features all higher-order corrections at the orders that we are interested in. This relation can be used to iteratively to determine the W-boson mass within the SM:
\begin{equation}
\mw^2=\mz^2(\frac{1}{2}+\sqrt{\frac{1}{4}-\frac{\alpha \pi}{\sqrt{2}G_{\mu}\mz^2}(1+\Delta r)})
\end{equation}\label{mw}

\subsubsection{Effective weak mixing angle $sin^2\theta_{eff}^f$}\label{sweff}
The effective weak mixing angle is defined as associated with the ratio of the Z-boson vector coupling form factor and the axial-vector coupling form factor. It is most sensitively determined at the Z-pole where the $Z/\gamma^*$ interference and photon exchange 
are suppressed. Hence we are interested in computing
\begin{equation}
sin^2\theta_{eff}^f=\frac{1}{4\lvert Q_f\rvert }(1+\Re{\frac{V_f(s)}{A_f(s)}})\Bigg\rvert_{s=\mz^2},
\end{equation}
where 
\begin{equation}\label{formfva}
\begin{split}
V_f(s)&=v_f^{Z}(s)-v_f^{\gamma}\frac{\Sigma_{\gamma Z}(s)}{s+\Sigma_{\gamma\gamma}(s)}\, \\
A_f(s)&=a_f^{Z}(s)-a_f^{\gamma}\frac{\Sigma_{\gamma Z}(s)}{s+\Sigma_{\gamma\gamma}(s)},\, \\
\end{split}
\end{equation}
where $v_f^X$ and $a_f^X$ are the effective vector and axial-vector couplings of vertices $Xf\bar{f}$, and the self-energy $\Sigma_{XY}$ stems from $\gamma-Z$ mixing at higher-orders.
%{\bf [I think you need to define $v_f^X$, $a_f^X$ and $\Sigma_{XY}$ here]}

\subsubsection{Partial Width $\overline{\Gamma}[Z\to f\bar{f}]$}\label{gamzff}
the partial width $\overline{\Gamma}$ can be recast by the Z-boson self-energy and vector/axial-vector couplings by applying optical theorem. It reads
\begin{equation}
\overline{\Gamma}_f = \frac{N_c^f\mz}{12\pi} C_\PZ \Bigl [
 {\cal R}_{\rm V}^f |V_f|^2 + {\cal R}_{\rm A}^f |A_f|^2 \Bigr ]_{s=\mz^2} 
 \;. \label{Gz}
\end{equation}
Here $N_c^f=3(1) $for quarks(leptons), and $C_\PZ$ is given by Z self-energy contributions at given orders. The radiators $\cal{R}_{\rm V,\rm A}$ contain final-state QCD and QED radiations. In our case, when only closed fermionic loops are considered, they are simply 1.

\section{Technical Aspects}
\label{sec:tech}
\begin{figure}[ht!]
\centering
\includegraphics[width=0.8\linewidth]{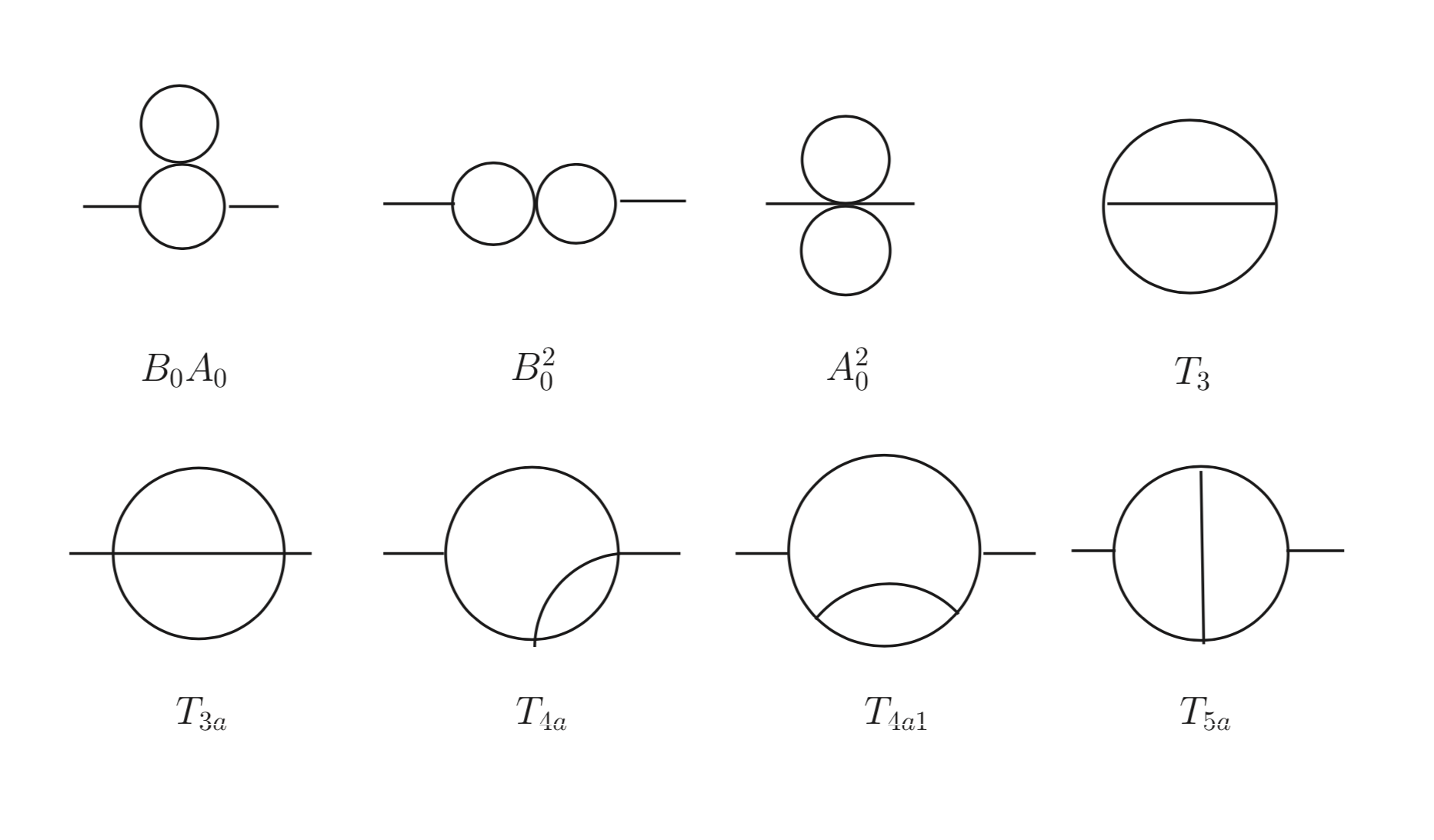}
\vspace{-1.5em}
\mycaption{The MI topologies used for genuine two-loop self-energy contributions, with notation taken from \cite{Bauberger:2019heh}.\label{fig:masint}}
\end{figure}
In this calculations we turned off CKM mixing and all fermion masses, except the top quark, due to their negligible numerical impact. {\sc FEYNARTS 3.3}\cite{feynarts} and {\sc FEYNCALC 9.2.0}\cite{feyncalc} are employed for amplitudes generation and algebraic reduction. The numerical evaluation is carried out by using {\sc TVID 2.0} \cite{Bauberger:2019heh}. Some $\mathcal{O}(D-4)$ coefficients from scalar one-loop integrals have been computed by following Eq.~4.1 in Ref.\cite{Nierste:1992wg}. When comparing with previous results with two fermionic loops in Refs.~\cite{mwshort,mwlong}, \cite{swlept} and
 \cite{gz}, we have found exact algebraic agreement except one term 
 \begin{equation}
 -\frac{d}{ds}\Big(\frac{\Bigl[\Im\Sigma_{\gamma Z(1)}(s)\bigr]^2}{s}\Big), \label{mis}
 \end{equation}
which stems from $\gamma-Z$ mixing at two-loop level in partial Z width, which was missing in Ref.~\cite{gz}. In Ref.~\cite{Chen:2020xzx}, this error has been corrected and its numerical impact was evaluated.
 
For genuine two-loop amplitudes, the MI reductions are done in two independent ways: integration-by-part (IBP) identities \cite{ibp} as implemented in {\sc FIRE6}\cite{fire6}, and the integral reduction techniques of Ref.~\cite{Weiglein:1993hd}. We should mention that, unlike one-loop cases, the choice of a MI basis at the two-loop level is not unique and may also not be minimal. One of the MI bases used in this calculation is shown in Fig~\ref{fig:masint}. However, despite the different choices of the MI basis, the two independent calculations by the authors agree numerically.
Furthermore, one must also compute the derivatives of two-loop self-energy functions to carry out the necessary renormalization counterterms. Care must be taken when deriving the derivative of the two-loop self-energy master integral with zero external momentum. With the help of chain rules, we obtain
\begin{equation}
\begin{split}
&\frac\partial{\partial p^2}I(...;p^2=0)=\frac{1}{2d}\frac{\partial^2}{\partial p_{\mu}\partial p^{\mu}}I(...;p^2)\bigg|_{p^2=0}\\
                     &\quad= \frac{2}{d}\biggl[\Bigl(1+a_2+a_5-\frac{d}{2}\Bigr)(a_2 {\bold 2^+}+a_5 {\bold 5^+})\\
                     &\qquad\quad+m_2^2 a_2(a_2+1){\bold 2^{++}}+m_5^2a_5(a_5+1){\bold 5^{++}}\\
                     &\qquad\quad +a_2a_5((m_2^2-m_3^2+m_5^2){\bold{2^+5^+}}-{\bold 2^+}{\bold3^-}{\bold 5^+})I
		     \biggr]_{p^2=0},
                     \end{split}                     
\end{equation}
whereas for $p^2 \neq 0$, one obtains \cite{gz}
\begin{equation}
\begin{split}
\frac{\partial}{\partial p^2} I(...;p^2\neq 0)  &=-\frac{1}{2p^2} p^{\mu} \frac{\partial}{\partial p^{\mu}} I(...; p^2)\\
                    & = -\frac{1}{2 p^2} \bigl[(a_2 +a_5) -a_2 {\bold1^-}{\bold 2^+}-a_5 {\bold 4^-}{\bold 5^+}\\
                    &\qquad\quad +a_2(m_2^2-m_1^2+p^2){\bold 2^+}+a_5(m_5^2-m_4^2+p^2){\bold 5^+}\bigr]I,
\end{split}
\end{equation}
where I is defined as the most generic two-loop self-energy master integral
\begin{equation}
\begin{aligned}
&I(a_1,a_2,..., m_1,m_2,..,; p^2) \\
&\equiv \int \frac{d^d q_1\,d^d q_2}{(q_1^2-m_1^2)^{a_1}((q_1+p)^2-m_2^2)^{a_2}((q_2-q_1)^2-m_3^2)^{a_3}(q_2^2-m_4^2)((q_2+p)^2-m_5^2)^{a_5}}\,
\end{aligned}
\end{equation}
and the standard lowering/raising operators are defined as
\begin{equation}
 {\bold 4^-}{\bold 5^+} I =I(a_4-1,a_5+1).
\end{equation}
Then one can apply IBP identities again to further reduce the raised/lowered MI integrals $I(...;p^2)$ down to the chosen MI basis such as Fig.~\ref{fig:masint}.

\section{Numerical Results}\label{sec:num}
%-------------------------------------------------------------
Given the benchmark inputs in Tab.~\ref{tab:input},
\begin{table}[tb]
\renewcommand{\arraystretch}{1.2}
\begin{center}
\begin{tabular}{|r@{$\;=\;$}ll|}
\hline
$\MZ$ & $91.1876\gev$ & {$\biggr\}\!\Rightarrow\;
 \mz = 91.1535\gev$} \\
$\Gamma_\PZ$ & $2.4952\gev$ & \\
$\MW$ & $80.358\gev$ & {$\biggr\}\!\Rightarrow\;
 \mw = 80.331\gev$} \\
$\Gamma_\PW$ & $2.089\gev$ & \\
$M_t$ & $173.0\gev$ & \\

$m_t(\mu=m_t)$ & $163.229 \gev. $& \\

$M_{f\neq \Pt}$ & 0 & \\
$\as$ & 0.1179 & \\
$\alpha$ & \multicolumn{2}{@{}l|}{$1/137.035999084$} \\
$\Delta\alpha$ & $0.05900$ & \\
$G_\mu$ & \multicolumn{2}{@{}l|}{$1.1663787 \times 10^{-5}\gev^{-2}$} \\
\hline
\end{tabular}
\end{center}
\vspace{-2ex}
\mycaption{Benchmark input parameters used in the numerical
analysis, based on Ref.~\cite{pdg}. Both benchmark values for alternative top-quark mass prescriptions are listed.
\label{tab:input}}
\end{table}
%%-------------------------------------------------------------
the numerical results for the leading fermionic contributions to all above-mentioned EWPOs at both $\mathcal{O}(\alpha^3)$ and mixed EW-QCD $\mathcal{O}(\alpha^2\alpha_s)$ are shown in Tab.~\ref{tab:numcomp}. It is evident that all the corrections computed at leading fermionic three-loop level are negligible for the precision tests conducted at the LEP and LHC, see Tab.~\ref{tab:futcoll}. However, one can also see that the experimental uncertainties mapped out by future $e^+e^-$ colliders are comparable to the three-loop corrections. Hence these corrections computed in Ref.~\cite{Chen:2020xzx} cannot be ignored. Combining the $\mathcal{O}(\alpha^3)$ and $\mathcal{O}(\alpha^2\alpha_s)$ corrections, we see $\Delta \mw$ and $\Delta'\overline{\Gamma}$ having a sizable corrections while others are subject to accidental cancellations.
 \begin{table}[tb]
\renewcommand{\arraystretch}{1.2}
\begin{tabular}{|l | r | r | r | r | r | r |}\hline
                     & $\Delta r $        & $\Delta \mw$ (MeV) & $\Delta \seff{}\;$ & $\Delta' \seff{}\;$ & {$\Delta\overline{\Gamma}_{\rm tot}$ [MeV]} & {$\Delta' \overline{\Gamma}_{\rm tot}$ [MeV]} \\ \hline
$\mathcal{O}(\alpha^3)$      &$2.5\times 10^{-5}$   & $-$0.389                     & $1.34\times 10^{-5}$   & $2.09 \times 10^{-5}$   & 0.331                                                  & 0.255                                                    \\ 
$\mathcal{O}(\alpha^2\as)$ &-0.000109& 1.703                   & $1.31\times 10^{-5}$   & $-1.98\times 10^{-5}$   & $-$0.103                                                 & 0.229                                                    \\ 
Sum                  & -0.000084     & 1.314                      & $2.65\times 10^{-5}$   & $0.11\times 10^{-5}$    & 0.228                                                  & 0.484                                                     \\ \hline
\end{tabular}
\mycaption{This table shows the numerical results of the leading fermionic three-loop corrections to EWPOs at $\mathcal{O}(\alpha^3)$ and at $\mathcal{O}(\alpha^2\as)$ from Ref.~\cite{Chen:2020xzx}. The EWPOs denoted with a prime use $M_W$ predicted from the Fermi constant $G_{\mu}$ rather than the value in Tab.~\ref{tab:input}. One can see that the two contributions have comparable size, except for $\Delta \overline{M}_W$, where the mixed EW--QCD three-loop correction is about four times larger in magnitude than the pure EW three-loop.} 
\label{tab:numcomp}
\end{table}
\begin{table}[tb]
\renewcommand{\arraystretch}{1.2}

\centering
\begin{tabular}{|l c| |c|| c | c | c |}\hline
                        &Global fits at LEP/SLD/LHC  &Current intrinsic theo. error      & CEPC & FCC-ee         & ILC/GigaZ \\
                         \hline
$\MW$[MeV]  &           $12$                            &$4(\alpha^3,\alpha^2\as)$ &  $1$        & $0.5\sim1$      & $2.5$   \\
$\Gamma_Z$[MeV]  &2.3       &$0.4(\alpha^3,\alpha^2\as,\alpha\as^2)$   & $0.5$       & $0.1$           &   $1.0$    \\
$\seff{f}$ [$10^{-5}$] &16&$4.5(\alpha^3,\alpha^2\as)$&  $2.3$      &          $0.6$               &             $1$  \\
\hline
\end{tabular}
\mycaption{This table demonstrates the current experimental uncertainties given by the global fits of measurements taken from the LEP, SLD, and LHC vs. future experimental
accuracies projected for CEPC, FCC-ee, and ILC for three EWPOs~\cite{cepc,fccee,Fujii:2019zll,Wilson:2016hne}. For ILC, the GigaZ option is considered, which is a $Z$-pole run with 100~fb$^{-1}$.}
\label{tab:futcoll}
\end{table}
When switching the top-quark mass from OS to {\msbar} prescription, using the benchmark value given in Tab.~\ref{tab:input},  the overall magnitude of leading fermionic $\OO(\alpha^2\as)$ corrections become noticeably smaller. This is normally expected as {\msbar} prescription converges faster than OS for QCD corrections. We perform the similar numerical evaluations summarized in Tab.~\ref{tab:msbarewpos}.

\begin{table}[ht]
\renewcommand{\arraystretch}{1.2}
\centering
\begin{tabular}[t]{|l|l|l|}
\hline
$X$ & $\Delta X_{(\alpha^2\as)}$ & $\Delta' X_{(\alpha^2\as)}$ \\
\hline
$\Delta r$ [$10^{-4}$] &$-$0.50& \\
$\Delta M_{\PW}$ [MeV] & \phantom{$-$}0.78&\\
$\seff{}$ [$10^{-5}$] & \phantom{$-$}0.75 & $-$0.76 \\
%$\Gamma_\ell$ [MeV] & $-$0.0003 & \phantom{$-$}0.0047 \\
%$\Gamma_\nu$ [MeV] & \phantom{$-$}0.0009 & \phantom{$-$}0.0086 \\
%$\Gamma_{\rm d}$ [MeV] & $-$0.0018 & \phantom{$-$}0.0223 \\
%$\Gamma_{\rm u}$ [MeV] & $-$0.0029 & \phantom{$-$}0.0183 \\
$\Gamma_{\rm tot}$ [MeV] & $-$0.0093 & \phantom{$-$}0.143 \\
\hline
\end{tabular}
\mycaption{Leading fermionic three-loop corrections to EWPOs at $\mathcal{O}(\alpha^2\as)$ with \msbar\ prescription for the top mass.}
\label{tab:msbarewpos}
\end{table}
A thorough comparison between OS and {\msbar} top mass prescription is given in Ref.~\cite{Chen:2020xzx} Tab.5, from which one observes that the numerical shifts at two-loop and three-loop levels partially compensate each other in both schemes as one would expect.

As mentioned above, a previous paper has missed the term \eqref{mis} contributing to $\Delta \overline{\Gamma}_f$ at two-loop order. This missing term results in numerical impact around $\OO(0.01)$~MeV to $\Delta \overline{\Gamma}_f$. This turns out to be relatively small but clearly non-negligible for the precision level we want to achieve at future colliders.

\section{Conclusions}
 In this proceeding, we highlight recent computations of leading fermionic three-loop corrections to EWPOs at both $\mathcal{O}(\alpha^3)$ and mixed EW-QCD $\mathcal{O}(\alpha^2\alpha_s)$. These computations are carried out in a fully gauge-invariant way. The numerical size of leading fermionic loop corrections should be considerably large due to the power of top mass and $N_f^n$ enhancement. However, they turn out to be milder than one would expect due to some accidental cancellations. Hence, other missing three-loop contributions may give corrections of similar magnitude, and they need to be included
 to further reduce the intrinsic theoretical uncertainty down to the level that matches the goals of future colliders.
%, we need to calculate other missing three-loop contributions, where a 
Here genuine electroweak three-loop integrals with various scales in the denominators will come into play, which will require significant additional work in the future.
\section*{Acknowledgements}
This work has been supported in part by the National science Foundation under grant no. PHY-1820760.

\bibliographystyle{SciPost_bibstyle} % Include this style file here only if you are not using our template
{100}
\end{document}